\documentclass[useAMS,usenatbib,usegraphicx]{mn2e}

\def\kms{km s$^{-1}$}
\def\msun{M$_{\sun}$}
\def\rsun{R$_{\sun}$}

\def\apjl{ApJ}
\def\apj{ApJ}
\def\apjs{ApJS}
\def\aj{AJ}
\def\mnras{MNRAS}

\def\apss{Ap\&SS}
\def\j0106{J0106$-$1000}

\title[The Shortest Period Binary WD]
{The Shortest Period Detached Binary White Dwarf System\thanks{Based
on observations obtained at the MMT Observatory, a joint facility of
the Smithsonian Institution and the University of Arizona.}}

\author[M. Kilic et al.]
       {Mukremin Kilic$^1$\thanks{\em Spitzer Fellow},
       Warren R. Brown$^1$,
       S. J. Kenyon$^1$,
       Carlos Allende Prieto$^{2,3}$, 
       \newauthor
       J. Andrews$^4$,
       S. J. Kleinman$^5$,
       K. I. Winget$^6$,
       D. E. Winget$^6$
       and J. J. Hermes$^6$\\
       $^1$Smithsonian Astrophysical Observatory, 60 Garden St, Cambridge, MA 02138, USA\\
       $^2$Instituto de Astrof\'{\i}sica de Canarias, E-38205 La Laguna, Tenerife, Spain\\
       $^3$Departamento de Astrof\'{\i}sica, Universidad de La Laguna, E-38206 La Laguna, Tenerife, Spain\\
       $^4$Columbia Astrophysics Laboratory, Columbia University, New York, NY 10027, USA\\
       $^5$Gemini Observatory, 670 N. A'ohoku Place, Hilo HI 96720, USA\\
       $^6$Dept. of Astronomy, University of Texas at Austin, RLM 16.236, Austin, TX 78712, USA
}

\begin{document}

\maketitle

\begin{abstract}

We identify SDSS J010657.39$-$100003.3 (hereafter \j0106) as the shortest period
detached binary white dwarf (WD) system currently known. We targeted \j0106 as part of our radial
velocity program to search for companions around known extremely low-mass (ELM,
$\sim$ 0.2\msun) WDs using the 6.5m MMT.
We detect peak-to-peak radial velocity variations of 740 \kms\ with an orbital period
of 39.1 min. The mass function and optical photometry rule out a main-sequence star companion.
Follow-up high-speed photometric observations obtained at the McDonald 2.1m telescope
reveal ellipsoidal variations from the distorted primary but no eclipses. 
This is the first example of a tidally distorted WD. Modeling the lightcurve,
we constrain the inclination angle of the system to be $67^{\circ} \pm 13^{\circ}$.
\j0106 contains a pair of WDs (0.17\msun\ primary + 0.43\msun\ invisible secondary)
at a separation of 0.32\rsun. The two WDs will merge in 37 Myr and most likely form
a core He-burning single subdwarf star. \j0106 is the shortest timescale merger system currently known.
The gravitational wave strain from \j0106 is at the detection limit of the
Laser Interferometer Space Antenna (LISA). However, accurate ephemeris and orbital period measurements
may enable LISA to detect \j0106 above the Galactic background noise.

\end{abstract}


\section{INTRODUCTION}

ELM WDs are ideal targets for finding binary WD merger systems.
Short period binary stars interact early in their stellar evolution,
experience enhanced mass-loss during one or two common-envelope phases \citep{sarna96}, and
end up as lower mass WDs. Thus the most compact binary systems are expected to form ELM WDs,
and a survey targeting ELM WDs should discover merging systems.

\citet{kilic09,kilic10,kilic11} and \citet{brown10} have established a radial velocity program,
the ELM Survey, to search for companions around known ELM WDs in the SDSS Data Release 7 footprint.
The discovery of 12 binary WD merger systems in a sample of
two dozen WDs observed to date has tripled the number of
known binary WD merger systems. Eighteen of the $M\leq0.25$\msun\ WDs in the ELM Survey
are in 1-24 hr period binaries with merger times as short as 100 Myr. Six
of these systems have extreme mass ratios ($M_1/M_2 = q \approx 0.2$), which may lead
to stable mass transfer AM CVn systems. If the mass-accreting
WDs in these systems are massive, they can potentially form Type Ia supernovae
\citep[SNe,][]{webbink84,iben84}. Alternatively, accretion of helium from a companion
may lead to the detonation of the surface helium layer on a C/O WD
in a fast and faint supernova, i.e. SNe ``.Ia'' \citep{bildsten07}.

Here we present the exciting discovery of a new binary system found in the ELM survey.
\j0106 was originally classified as a subdwarf star in the SDSS DR4 WD catalog of
\citet{eisenstein06}. \citet{kleinman10} re-classified it as an ELM WD based on
a reanalysis of the SDSS spectroscopy. Our follow-up radial velocity and high
speed photometric observations demonstrate that \j0106 contains a pair of WDs with
an orbital period of only 39.1 minutes. This system, the shortest period detached binary
WD system known, presents the first detection of a tidally distorted WD.

In Section 2 we describe our spectroscopic and photometric observations. 
In Sections 3 and 4 we constrain the physical parameters of the binary and discuss
the nature and future evolution of the \j0106 system. We conclude in Section 5.

\section{OBSERVATIONS}

We used the 6.5m MMT with the Blue Channel spectrograph to obtain medium resolution spectroscopy
of \j0106 on UT 2010 Dec $1-3$. 
We operate the spectrograph with the 832 line mm$^{-1}$
grating in second order, providing wavelength coverage from 3600 \AA\ to 4500 \AA\ and a spectral
resolution of 1.2 \AA. We obtain all observations at the parallactic angle, with a comparison
lamp exposure paired with every observation. We flux-calibrate using blue
spectrophotometric standards \citep{massey88}, and we measure radial velocities
using the cross-correlation package RVSAO. The details of our data
reduction procedures are discussed in \citet{kilic09,kilic10}. We check the stability of the
spectrograph using the Hg line at 4358.34 \AA\ from Tucson/Nogales street lights.
We measure an average velocity offset of $-0.9 \pm 0.3$ \kms\ for this line over
the entire three nights of observations.

\j0106 is relatively faint ($g=19.8$ mag). We started our observations
with 10 min exposures. After detecting $>400$ \kms\ velocity variations in 15
min, we decreased the individual exposure times to 8 min.
Realizing that \j0106 is a very short period system after our second night of observing, we
also acquired high speed photometric observations of \j0106 using the McDonald 2.1m
Otto Struve Telescope with the Argos frame transfer camera \citep{mukadam05} on UT 2010 Dec 3. 
Argos provides a field of view of $2.8' \times 2.8'$.
We obtained time series photometry of \j0106 with the BG40 filter every 30 s for about 2.6 hr.
The Argos field-of-view includes several comparison stars that are useful for relative photometry.

\section{RESULTS}

\subsection{The Orbital Period}

Table 1 lists our radial velocity measurements for \j0106.
We compute best-fit orbital elements using the code of \citet{kenyon86},
which weights each velocity measurement by its associated error. The uncertainties
in the orbital elements are derived from the covariance matrix and $\chi^2$. To
verify these uncertainty estimates, we perform a Monte Carlo analysis using
$10^4$ sets of simulated radial velocities. We adopt the inter-quartile range in the
period and orbital elements as the uncertainty.

\begin{table}
\centering
\caption{Radial Velocity Measurements for \j0106}
\begin{tabular}{cr}
\hline 
HJD$-$2455530 & $v_{helio}$ \\
(days) & (\kms) \\
\hline
1.63742 &    313.3 $\pm$ 26.0 \\
1.69330 &    330.1 $\pm$ 24.5 \\
1.74373 &    116.4 $\pm$ 37.5 \\
2.65076 &    127.9 $\pm$ 36.2 \\
2.66124 & $-$292.6 $\pm$  8.8 \\
2.67089 &    329.7 $\pm$ 13.0 \\
2.67802 &     32.4 $\pm$ 16.7 \\
2.68510 & $-$362.4 $\pm$ 13.0 \\
2.69110 & $-$146.4 $\pm$ 16.7 \\
2.69707 &    320.9 $\pm$ 21.8 \\
2.70401 &    200.7 $\pm$ 18.7 \\
2.70998 & $-$297.8 $\pm$ 28.9 \\
2.71774 & $-$132.5 $\pm$ 22.7 \\
2.72292 &    219.2 $\pm$ 27.7 \\
2.72909 &    366.8 $\pm$ 26.2 \\
3.63764 & $-$365.2 $\pm$ 15.0 \\
3.64378 &     51.2 $\pm$ 17.7 \\
3.65060 &    384.8 $\pm$ 21.0 \\
3.65659 &      9.3 $\pm$ 15.0 \\
3.66333 & $-$402.0 $\pm$ 40.2 \\
3.67095 &    109.9 $\pm$ 11.6 \\
\hline
\end{tabular}
\end{table}

\begin{figure}
\includegraphics[width=2.6in,angle=-90]{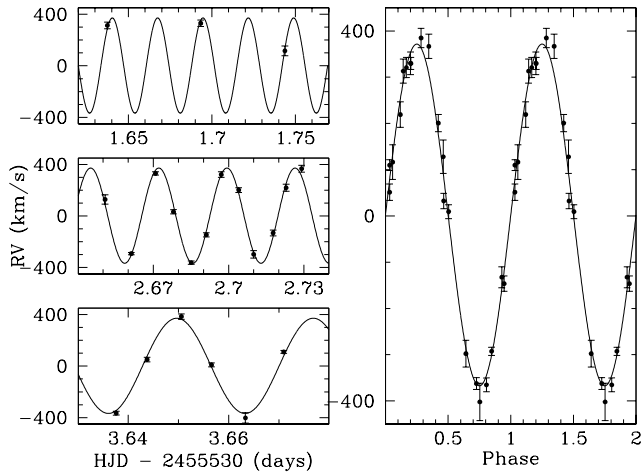}
\caption{The radial velocities of \j0106 observed over three nights in 2010 December
(left panels). The right panel shows all of these data points phased with the best-fit
period. The solid line represents the best-fit model for a circular orbit with a period
of 39.1 min and $K=369.5$ \kms.}
\end{figure}   

\j0106 exhibits radial velocity variations with a semi-amplitude of $K = 369.5 \pm 3.6$
\kms and orbital period of $P = 0.027153 \pm 0.0000195$ d, or 39.100 $\pm$ 0.028 min.
Figure 1 shows the best-fit orbit compared to the observed radial velocities.
The relatively long exposure times (8 min) compared to the orbital period (39.1 min) results
in an underestimated velocity semi-amplitude, $K$. This is a direct consequence of the sine curve 
not being linear when the velocities are at the extremes. To verify this effect, we sampled a sine
curve at the exact 21 phases of our observations with $P/5$ long integrations. We recover the exact
period, but $K$ is systematically underestimated by 6.5\%. Thus, the corrected velocity semi-amplitude for \j0106
is $K= 395.2$ \kms. With this correction,
\j0106 has a mass function of $f= 0.1736 \pm 0.0047$ \msun.
The systemic velocity (after subtracting the gravitational redshift of 1.9 \kms) is 0.3 $\pm$ 2.7
\kms\ and the time of spectroscopic conjunction is HJD 2455531.633574 $\pm$ 0.000129 d.

\subsection{The Physical Parameters of the Binary}

Our time-series spectroscopy provides for robust determinations of effective
temperature and surface gravity. We perform stellar atmosphere model fits using
synthetic DA WD spectra kindly provided by D.\ Koester. The grid of WD model
atmospheres covers effective temperatures from 6000 K to 30,000 K in steps of
500 K to 2000 K, and surface gravities from $\log{g}=$ 5.0 to 9.0 in steps of
0.25 dex. We perform fits to the Balmer line profiles using the average composite
spectra. We also perform fits to the individual spectra to derive a robust statistical
error estimate. Figure 2 shows the observed Balmer line profiles (jagged lines) for \j0106
compared to our best-fit model (solid line). The best-fit model has
$T_{\rm eff} = 16485 \pm 456$ K and $\log{g} = 6.01 \pm 0.04$. 

Previous stellar atmosphere fits find a comparable $\log g$ but a 2500 K hotter $T_{\rm eff}$
\citep{kleinman10}. The discrepancy in their temperature estimate is most likely due to the
low signal-to-noise SDSS spectrum. Our best-fit model matches our S/N = 40 spectrum and the
SDSS photometry (Figure 2).

\begin{figure}
\includegraphics[width=3.4in]{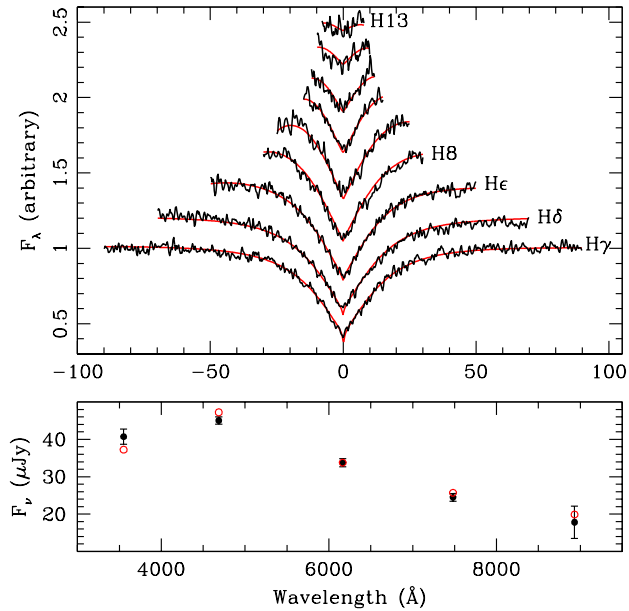}
\caption{Model fits (red lines) to the Balmer line profiles of \j0106 (jagged lines, top panel).
The spectral energy distribution of \j0106 (based on the SDSS photometry, filled circles)
compared to the best-fit model (open circles, bottom panel).}
\end{figure}

Based on the improved \citet{panei07} tracks \citep[see][]{kilic10} for ELM WDs,
\j0106 is a 1.1 Gyr old\footnote{This age estimate is somewhat uncertain due to the assumption on
the thickness of the surface hydrogen layer for 0.17\msun WDs.}
 0.17\msun\ WD with a radius $R=$ 0.057 \rsun. Its absolute
magnitude $M_g = 7.8$ corresponds to a distance of 2.4 kpc. Based on five epochs from
the USNO-B and the SDSS, \citet{munn04} measure a proper motion of 
($\mu_{\alpha} cos \delta, \mu_{\delta}) = (20.2, -10.5$) mas yr$^{-1}$.
\j0106 is 2.3 kpc below the Galactic plane and it has $U=-115 \pm 43, V=-222 \pm 43$,
and $W=-15 \pm 12$ km s$^{-1}$ with respect to the local standard of rest \citep{hogg05}.
Clearly, \j0106 is a halo star.

The mass function implies a companion mass of $\geq$0.37 \msun.
The orbit is far too small to contain a main-sequence star of 0.37 \msun\ or more .
Therefore, the companion is a compact object. Based on the mass function alone,
the probability of a neutron star (1.4-3.0 \msun) companion is 7.3\%. The probability
of a SNe Ia, for which the companion would be a $1.23-1.40$ \msun\ WD, is only 1.9\%. However, if
sub-Chandrasekhar mass WDs do explode as Type Ia SNe \citep{vankerkwijk10}, this
probability may be higher.

\subsection{The Light Curve}

The chance of an eclipse is relatively high for \j0106. For an edge-on orbit,
the companion would be a 0.37\msun\ WD. To avoid detection in the SDSS photometry,
we assume that the companion is 10$\times$ fainter than the visible WD and thus it has
$T_{\rm eff} \leq 14600$ K and $R\approx0.02$\rsun.
The eclipse depth and duration would be 12\% and $\approx 90$ s (assuming a total eclipse), respectively.
Due to the relatively large size of the visible WD compared to the orbital
separation, the probability of a grazing eclipse is 25\%. 

Figure 3 shows the Argos light curve of \j0106 (top panel) over four orbits.
\j0106 is relatively faint and the Argos light curve has a few percent scatter.
We do not detect any pulsations at the $\geq0.8$\% level.
However,
the fourier transform of the \j0106 light curve reveals a significant peak at 1187
$\pm$ 13 s (half the orbital period) with an amplitude of 1.7\% $\pm$ 0.3\%.
We use the ephemeris from the radial velocity observations obtained on the same
night to calculate the phase for our photometric observations. The middle panel
in Figure 3 shows the light curve folded over the best-fit orbital period. There
are essentially four separate observations every 30 s in the folded light curve.
The bottom panel in Figure 3 shows the same light curve binned by four points.
This panel
clearly shows the ellipsoidal variations; the ELM WD
is distorted due to the companion. This is the first detection of ellipsoidal variations
for a WD.

\begin{figure}
\includegraphics[width=3.4in]{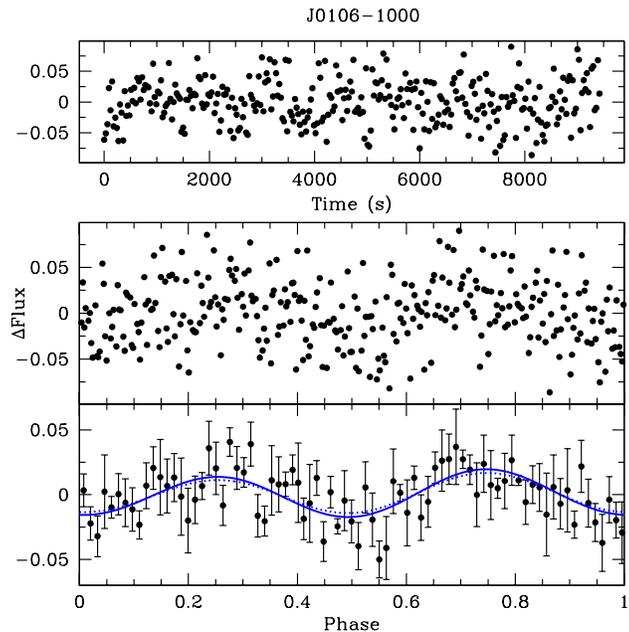}
\caption{High speed photometry of \j0106 over four orbital periods (top
panel). The middle and bottom panels show the same light curve folded over
the orbital period and binned by four points, respectively.
Our best-fit model ($i=73^{\circ}$) and a comparable model ($i=60^{\circ}$) are shown
as solid and dotted lines.}
\end{figure}

To verify that the observed variations in the light curve are not a statistical fluctuation,
we perform a bootstrap analysis. We randomly permute the light curve points in time and
create $10^5$ simulated light curves. This analysis shows that the probability of the measured
ellipsoidal variation signal being a random one is smaller than $10^{-5}$. We also checked
for variations in the brightest reference star by folding its light curve for the
39.1 min period of \j0106. This analysis did not reveal any periodic variations in the reference
star. These tests show that the detected variability is highly significant.

To model the ellipsoidal variations and the reflection effect, we follow \citet{morris93}.
In their formalism, the amplitude of the ellipsoidal effect is roughly
$\delta f_{ell} = (m_2/m_1)(r_1/a)^3$, where $a$ is the orbital semi-major axis and $r_1$
is the radius of the primary \citep{zucker07,shporer10}. For \j0106,
$(m_2/m_1) \approx 2.2$ for an edge-on orbit, yielding $\delta f_{ell}=$ 1.4\%.
For a more accurate estimate, we follow equation 1 in \citet{morris93} and include
the low order terms up to the cos 4$\phi$ term, where
$\phi = 0^{\circ}$ when the primary star is farthest from the observer.
The transmission curve of the Argos camera with the BG40 filter is similar to a $B$-band filter.
We use limb darkening and gravity-darkening coefficients of $u_1 = 0.36$ and $\tau_1=0.487$, respectively
\citep[see equation 1 and Table 1 in][]{morris93}. The results change only slightly for different 
limb-darkening and gravity-darkening coefficients.
The ellipsoidal variations are dominated by the
cos 2$\phi$ term, which has an amplitude of 1.8\% for the \j0106 system at
90$^{\circ}$ inclination.
We estimate the contribution from the reflection effect to be $<$0.1\% and
therefore neglect it. On the other hand, the relativistic beaming effect is
expected to be about 0.3\% \citep{maxted00,shporer10}. We include the beaming effect,
but not the eclipses in our calculations. 

The amplitudes of the ellipsoidal variations and the relativistic beaming effect depend
on the companion mass and orbital separation, which depend on the inclination of the system.
Hence, the predicted amplitudes can be represented by a single parameter, the inclination angle.
We create model light curves for each inclination angle and chose the model with
the least $\chi^2$ as the best-fit model (shown as a solid line in Figure 3). 
This model has an inclination angle of $i=73^{\circ}$ and it matches the phase
and amplitude of the variations relatively well. The absence of eclipses
in the photometry is also consistent with this inclination angle.
The dotted line shows a comparable model with an inclination angle of $i=60^{\circ}$.
The similarities between
the two models and the relatively large scatter in our photometry indicate that
the error in inclination is relatively large.

To constrain the inclination angle of the system more accurately, we perform a Monte Carlo
analysis where we replace the measured flux $f$ with $f + g~\delta f$, where $\delta f$ is
the error in flux and $g$ is a Gaussian deviate with zero mean and unit variance. For each
of the $10^5$ sets of simulated light curves, we repeat our analysis and find the best-fitting
model (and the inclination angle). We adopt the interquartile range as the
uncertainty. The Monte Carlo analysis shows that the \j0106 light curve is best explained by a model
with $i = 67^{\circ} \pm 13^{\circ}$. Therefore, the companion is most likely a 0.43\msun\ object
at an orbital separation of 0.32\rsun. The mass ratio of the binary is $q=0.4$ and the merger time
due to gravitational wave radiation is 37 Myr.

\section{DISCUSSION}

\j0106 is the shortest period detached binary WD system currently known. Even though its orbital
period is comparable to the AM CVn systems, its spectrum shows only hydrogen absorption lines.
There is no evidence
of interaction or mass accretion between the two components other than the slight distortion of
the ELM WD due to mutual gravitation. The observed ellipsoidal variations are extremely useful
for constraining the inclination angle and the masses of both components of the system. The optical
spectroscopy and photometry of this system is best explained by a binary system containing two He-core WDs
with $M_1=$ 0.17\msun\ and $M_2=$ 0.43\msun\ at a separation of 0.32\rsun.

The future evolution of the system depends on the mass ratio of the two components. For a mass
ratio of $q=0.4$, \j0106 will likely have unstable mass transfer and merge. \citet{dan11} simulate
the mergers of double degenerate systems including 0.2\msun\ ELM WDs with 0.3-0.8\msun\ WD companions.
Their smoothed-particle hydrodynamic models indicate that systems with $q\geq0.25$ have
unstable mass transfer and merge after 20-80 orbits \citep[although see][for the possibility of stable
mass transfer]{motl07,racine07}. Unstable mass transfer may also lead to the detonation of the surface
helium layer on a C/O WD via Kelvin-Helmholtz instabilities. \citet{guillochon10}
find that the required conditions for triggering a surface explosion are only achieved in $\geq0.8$\msun\ WD
accretors with $>$0.2\msun\ companions. Hence, no efficient carbon burning is expected in binary WD systems
containing 0.45-0.8\msun\ C/O WD accretors and ELM WD donors. Based on these results, \j0106 will likely
merge and create a 0.6\msun\ core-He burning subdwarf in 37 Myr. This mass is close to the
canonical mass of 0.5\msun\ for subdwarfs \citep{heber09}.

\citet{brown11} estimate the merger rate of ELM WDs from a complete, color-selected sample of ELM WDs
found in the Hypervelocity Star Survey \citep{brown06,brown09}. Roughly 5 $\times 10^4$
(with a factor of few uncertainty) ELM WDs formed in the
Galactic disk in the last Gyr. About 70\% of these systems merge in less than a Gyr, and 66\% of these
systems have mass ratios $q\geq$0.25. From these estimates, roughly 2.3 $\times 10^4$ ELM WDs merged in the past Gyr.
There are three stars with $q\geq$0.25 and merger times shorter than a Hubble time in their sample. These three
stars, J0818+3536, J0923+3028, and J1053+5200, contain 0.17-0.23\msun\ + 0.33-0.44\msun\ WDs (assuming an average
inclination angle of 60$^{\circ}$). Based on the evolutionary calculations by \citet{dan11}, they are likely
to form 0.5-0.67\msun\ single subdwarfs. Hence, the formation rate of single subdwarfs through
mergers of ELM WDs is roughly 2.3 $\times 10^4$ in the last Gyr. Adding \j0106 to this sample would increase
this rate by about 50\% due to its relatively short merger time.

\citet{nelemans10} presents population synthesis models for
the Galactic population of subdwarf B stars. He predicts a total number of 5.6 $\times 10^5$ single subdwarfs
to form as a result of He WD mergers. The resulting mass distribution is centered around 0.5\msun\ with
a tail toward higher masses, similar to the mass distribution of the merger systems discussed above. Hence, the
ELM WD merger systems contribute significantly to the population of single subdwarfs in the Galaxy.

Short period binary WDs are important gravitational wave sources. \citet{nelemans04} and \citet{roelofs07} argue that
about half a dozen AM CVn binaries should be detected by LISA. With an orbital
period similar to the known AM CVn systems, \j0106 may be a promising candidate for detection. The orbital period,
inclination, and model-dependent distance estimate for \j0106 yield the gravitational wave strain at Earth,
$\log h = -22.7$ at a frequency $\log \nu$ (Hz) = $-3.07$ \citep{roelofs07}. This is at the S/N = 1
detection limit of LISA after 1 year of observations. Confusion with Galactic noise sources decreases
with accurately known orbital periods. Our ephemeris and orbital period measurements may enable LISA
to detect \j0106 above the Galactic background noise after a few years of observations.

\section{CONCLUSIONS}

We discovered the shortest period detached binary WD system currently known.
This system also presents the first detection of a tidally distorted WD.
We constrain the inclination angle of the system using high-speed photometric observations.
\j0106 contains a pair of low-mass WDs at an inclination angle of
$67^{\circ} \pm 13^{\circ}$. Follow-up high-speed photometric observations at a larger telescope
will be useful to better constrain the inclination (and therefore the companion mass) and to search
for grazing eclipses. The two WDs will merge in 37 Myr and most likely form a core He-burning single
subdwarf star.

\section*{Acknowledgements}

We thank A. Shporer and S. Nissanke for stimulating discussions,
D. Koester for kindly providing WD model spectra, and an anonymous
referee for useful suggestions.
Support for this work was provided by NASA through the {\em Spitzer Space Telescope}
Fellowship Program, under an award from Caltech.
KIW, DEW, and JJH gratefully acknowledge
the support of the NSF under grant AST-0909107 and the Norman
Hackerman Advanced Research Program under grants 003658-0255-2007 and 
003658-0252-2009.

\end{document}